\documentclass[conference]{IEEEtran}

\usepackage{amsmath,amssymb,amsthm}
\newtheorem{theorem}{Theorem}
\newtheorem{corollary}{Corollary}
\newtheorem{lemma}{Lemma}
\usepackage{graphicx}
\usepackage{booktabs}
\usepackage{multirow}
\usepackage{xcolor}
\usepackage{pifont}
\usepackage{url}
\usepackage{microtype}
\usepackage[hidelinks]{hyperref}

\definecolor{pend}{HTML}{C0561B}

\newcommand{\topk}{\mathrm{Top}}
\newcommand{\cov}{\mathrm{cov}}
\newcommand{\debt}{\mathrm{debt}}
\newcommand{\Cserve}{C_{\mathrm{serve}}}



\begin{document}


\title{[Experiment, Analysis, and Benchmark]\\When the Bill Falls: Reusing the Old Index Defers,\\Never Avoids, Re-encoding in Complete Embedding-Model Migration}


\author{
\IEEEauthorblockN{
Nannan Zhao,
Jiameng Zhang,
Taoyu Zhong,
Hao Wang,
and Bo Wang
}
\IEEEauthorblockA{
\textit{Northwestern Polytechnical University}\\
Xi'an, China\\
nannanzhao@nwpu.edu.cn,
\{zhangjiameng, zhongtaoyu, kingshulk989, wb2022302586\}@mail.nwpu.edu.cn
}
\thanks{Corresponding author: Nannan Zhao (nannanzhao@nwpu.edu.cn).}
}

\maketitle

\begin{abstract}
Upgrading the embedding model underlying a vector index requires re-encoding the entire corpus.
Recent literature proposes reusing the old index through several strategies: adapters translate
new-model queries into the old embedding space; window re-ranking lets the old index nominate
candidates for the new model to re-score while new vectors are encoded in the background. Testing
them on sixteen model pairs over a passage and a code corpus, and eleven more on Natural Questions
(NQ), we find three things. First, reuse never avoids re-encoding: once the migration must finish,
every strategy encodes every document, and reuse changes only when that work is paid and how much
quality is lost in the meantime. Second, adapters are evaluated against the wrong baseline: what
matters is whether they beat the old model, and of five adapter families only bridges between two
large language model (LLM) embedders of the same family do so, by a margin too small to count as an improvement. Third,
the cost of window re-ranking at any depth can be computed from the old index alone, under a bound
we prove for every workload; pre-registered replays on the full 8,841,823-passage collection
confirmed this prediction on all three pairs tested. Workload skew can also change the best
strategy: across 36 further pre-registered Zipf-skewed replays, the planned strategy held on 32.
From these findings we derive three rules, all computed before migrating: whether reuse is
worthwhile, which strategy to use, and how deep a window to take. The depth rule, committed before
its replays, priced every window that reached target quality to within 4 parts in $10^5$ of the
corpus. On NQ, a plan committed before any replay chose the strategy on 10 of 11 pairs and priced
every window exactly. The other two rules agree with the 13 core replays, the second among the
strategies it considers. The quality inputs, the gap between the models and the window's quality loss, transferred less
completely: they must be measured on the index that will serve.
\end{abstract}


\section{Introduction}\label{sec:intro}
Dense retrieval ties an index to a single embedding model. The standard practice for an upgrade
is a blue--green rebuild: re-encode the corpus with the new model, build a second index, validate
it, switch traffic~\cite{zilliz-migration}, and retire the old index. Until it is retired, the old
index doubles storage and every write path, and retiring it requires the migration to finish; at
hundreds of millions of documents, that takes days of accelerator time. Recent literature proposes three ways to reuse
the old index instead. Adapters learn a map that translates new-model queries into the old
embedding space, so that the old index can keep serving~\cite{drift-adapter}; they aim to avoid
re-encoding documents. Window re-ranking lets the old index nominate candidates, re-scores them
with the new model, and encodes new vectors as queries arrive and in the background~\cite{embedflow};
it defers encoding. Structure reuse exploits the old graph index to build the new one
faster~\cite{dgm}; it neither avoids nor defers encoding.

\textbf{Observations.} We tested these methods on sixteen model pairs over a
$10^6$-passage MS MARCO~\cite{msmarco} corpus and a code corpus, and on eleven more pairs on
NQ~\cite{nq}, asking what each
method saves and how much quality it loses. Consider a typical pair (P2, Table~\ref{tab:pairs}): an index
served by MiniLM~\cite{minilm} and upgraded to bge-base~\cite{bge}, which raises Recall@10 from 0.8335 to 0.8805.
(i)~\emph{Adapters are judged against the wrong baseline.} The literature reports adapter recall
relative to the new model, with Drift-Adapter reaching 95--99\%~\cite{drift-adapter}. The decision,
however, asks whether an adapter beats the old model, since an adapter that serves below it is
worse than not migrating. On this pair, that requires 94.7\% of the new model's recall, and the
smaller the upgrade, the closer the bar is to 100\%. In our replays (each a deterministic re-execution of a strategy
over a fixed arrival stream), a Procrustes adapter~\cite{procrustes}, a learned rotation between
the two spaces, reaches 46.8\%, and the best
of five adapter families 64\%. Only on the P5r pair
(Qwen3-Embedding-0.6B$\to$8B~\cite{qwen3emb}; Table~\ref{tab:adapters}) does any family clear the old model, by a
margin too small to count as an improvement; on every other pair every family serves below it, the
replayed one by up to 0.7843 Recall@10.
(ii)~\emph{Reuse does not save encoding.} If the old index is to be retired, every strategy
eventually encodes every document; reuse changes only when the work is paid. On this pair, by the
time the evaluated queries are served, window re-ranking has re-encoded 26.4\% of the
corpus and serves within 0.0023 Recall@10 of the new model, while the rebuild's
background has encoded only 3\%. None of the cited work prices this cost in advance.

\textbf{Three decisions.} Together with the fact that structure reuse
neither avoids nor defers encoding, these observations split the migration decision into three
questions, asked in order (Section~\ref{sec:plan}). First, is the upgrade worth doing? While a
rebuild runs, the old model keeps serving, so the quality lost is the gap between the two models;
if that gap is within a tolerance declared in advance, the upgrade buys nothing distinguishable,
and if it goes ahead a plain rebuild loses nothing either. Second, if it is
worth doing, which strategy should be used? By~(i), an adapter is considered only if it beats the
old model; structure reuse is not an alternative; what remains is the window. Third, how
deep should the window be? By~(ii), a window defers encoding, so the deferral must be priced, and
the old index alone suffices to price it: counting how many queries' windows cover each document
predicts how many documents will have been encoded after any number of queries, without encoding
any of them (Section~\ref{sec:model}), and an independent-window form bounds this cost for every
workload. Like a column histogram in a database optimizer, it prices a plan without executing
it. A deeper window costs more and loses less quality, which is measured on a judged sample.

\begin{figure}[htbp]
\centering
\includegraphics[width=0.88\columnwidth]{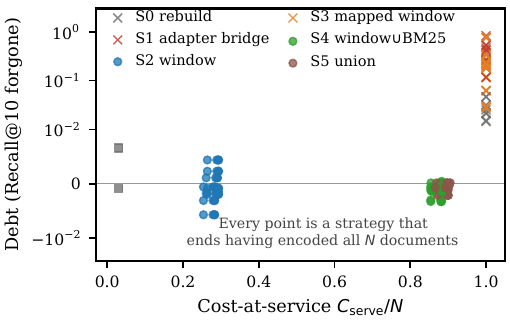}
\caption{Decision space at the primary operating point, ten text pairs. Every strategy ends having
encoded all $N$ documents; they differ in when the work is paid and in debt. $\circ$ work done;
$\square$ rebuild at target quality on background progress alone; $\times$ never at target
quality (charged $N$).}
\label{fig:space}
\end{figure}

\textbf{Tested as predictions.} Fig.~\ref{fig:space} shows the decision space. On the thirteen
core pairs, the first two rules can only be checked for consistency, because those replays predate
the planner. The first rule agrees with every replay (13 of 13; all
3 of its negatives have gaps within $10^{-3}$ of zero). The second picks the replay's
winner among the strategies it considers on all 10 migrating pairs (and on
45 of 45 seeded replays), and the winner among all six strategies on 9; on the
tenth, a lexical union beat the window by more than the tolerance on every seed. The genuine
predictions are those of the third rule. Committed before its replays, the 25-document window was
priced to within 4 parts in $10^5$ of the corpus on every pair where it reached
target quality (7 of 8), and its served quality matched the static estimate
within the tolerance on 7 of 8. On NQ, a plan committed before any replay
chose the strategy on 10 of 11 pairs and priced every window exactly.{}
On the full 8,841,823 passages, the coverage predictions held, but the quality inputs measured at
$10^6$ did not carry over; measured at the serving size, they predicted all six pre-registered
replays at 2M and 8.8M (Section~\ref{sec:scale}).{} Table~\ref{tab:claims} lists each finding
with its evidence.

\begin{table}[t]
\centering
\caption{Findings and their evidence. Proven: Section~\ref{sec:model}; predicted: committed before
the measurement that tests it; retrospective: consistent with the replay, not predicted.}
\label{tab:claims}
\footnotesize\setlength{\tabcolsep}{2.5pt}
\begin{tabular}{@{}p{0.40\columnwidth}p{0.43\columnwidth}l@{}}
\toprule
Finding & Evidence & Where \\
\midrule
Reuse defers, never avoids, re-encoding & by construction; spread measured & Fig.~\ref{fig:space} \\
Bridges must beat the old model & measured: 5 families $\times$ 15 pairs & Tab.~\ref{tab:adapters} \\
Window cost = coverage of served windows & measured: every window replay & Tab.~\ref{tab:econ} \\
Independent-window bound & proven; 1,237 uniform{} + 2,088 Zipf{} points; held at 8.8M& Fig.~\ref{fig:curve} \\
Plain window is the default; unions buy quality at 3.2--5.7$\times$ the cost; no selector beats a constant & measured; 5 pre-registered hypotheses; 45 seeded replays& Tab.~\ref{tab:hyp} \\
Skew shrinks the judged sample & predicted: 36 Zipf replays, family 32 of 36; 4 earlier & \S\ref{sec:setup} \\
D1/D2: the family & retrospective: 12 of 13& Tab.~\ref{tab:plan} \\
D3: the depth and its price & predicted: $K'{=}25$; NQ, 11 pairs & Tab.~\ref{tab:d3} \\
Gap and $G$ at $10^6$ do not bound them at 8.8M & pre-registered bound, falsified on 1 of 3 pairs & Tab.~\ref{tab:scale} \\
Window cost, quality and family at 2M and 8.8M & predicted: 6 replays, all hold & \S\ref{sec:scale} \\ \midrule
\multicolumn{3}{@{}p{0.97\columnwidth}@{}}{\emph{Not covered:} serving under load; scale replay beyond three pairs at $K'{=}50$; other GPUs (only the
latency numbers depend on hardware); languages other than English; long documents;
deliberately incomplete migration.}\\
\bottomrule
\end{tabular}
\end{table}


\section{Problem Setting and Evaluation Frame}\label{sec:frame}
Before comparing strategies, we must fix three things: what we count, when we measure quality,
and which strategies may win. Otherwise, each strategy could be charged in the way that flatters
it: a window strategy could be scored on documents it has not yet encoded, and a rebuild could be
measured only after its background has finished. We therefore count document encodings under the
new model, measure recall when each query is first served, and let a strategy win only if it
completes, reaches target quality, and serves no worse than the old model.
Section~\ref{sec:quant} then defines the four quantities used throughout.

\subsection{Setting and strategies}\label{sec:strategies}
Let $D$ be a corpus of $N$ documents, $f_s$ the source (old) encoder and $f_t$ the target (new)
encoder. An index $I_s$ over $\{f_s(d)\}_{d\in D}$ is serving. A workload $Q$ arrives over
time, possibly with repeats. $\topk^t_K(q)$ is the native top-$K$ of $q$ under $f_t$ over all
of $D$, and $\topk^s_{K'}(q)$, the top-$K'$ returned by $I_s$, is the \emph{window}: the
candidates the old index nominates for $q$. We use $K{=}10$ throughout. Quality is Recall@10,
the share of a query's judged relevant documents in its top 10: $M_S$ and $M_T$ for the source
and target models each searching the full corpus exactly, and $M_A$ for an adapter bridge (S1).

Table~\ref{tab:strategies} lists the six strategies. \emph{S0} (blue--green rebuild) serves the
old index while the corpus is re-encoded in the background, then switches. \emph{S1} (adapter
bridge~\cite{drift-adapter}) maps each new-model query into the old space with a learned map $A$
and lets the old index answer. \emph{S2} (window re-ranking~\cite{embedflow}) lets the old index
nominate $K'$ candidates, encodes inline those not yet encoded, and re-scores them with $f_t$.
\emph{S3} nominates with the bridged query $A(f_t(q))$ instead; \emph{S4} adds the
BM25~\cite{bm25} candidates a hybrid system already computes, free to compute but each to be
encoded under $f_t$ before it can be re-scored; \emph{S5} unions all three. We call S1 and S3 the \emph{mapping family} and S2, S4 and S5 the
\emph{window family}. Every window strategy
also searches a small target-space index over migrated and newly written documents, which have
no source vector. S1--S5 do not by themselves encode the whole corpus, so each runs with a
background re-encoder in one of five orders: \emph{rand}; \emph{size} (shortest first);
\emph{demand} (only documents surfaced by queries, which need never complete and is therefore
inadmissible); \emph{demand-then-rand}, which completes; and \emph{signal}, a document-level
priority (Section~\ref{sec:res-select}).

\begin{table}[t]
\centering
\caption{The six strategies. S1 and S3 form the mapping family; S2, S4 and S5 the window family.}
\label{tab:strategies}
\small\setlength{\tabcolsep}{4pt}
\begin{tabular}{@{}lll@{}}
\toprule
ID & Nominator & Ranker \\
\midrule
S0 rebuild & old index until switch & $f_s$, then $f_t$ \\
S1 adapter bridge~\cite{drift-adapter} & $I_s$ via $A(f_t(q))$ & $f_s$ \\
S2 window re-rank~\cite{embedflow} & $I_s$ via $f_s(q)$ & $f_t$ \\
S3 mapped window & $I_s$ via $A(f_t(q))$ & $f_t$ \\
S4 window $\cup$ lexical & $I_s\,\cup$ BM25~\cite{bm25} & $f_t$ \\
S5 window $\cup$ mapped $\cup$ lexical & all three & $f_t$ \\
\bottomrule
\end{tabular}
\end{table}

\subsection{What we count: document encodings}
The dominant cost of migration is encoding documents with $f_t$, so we count document
encodings, and \emph{total encoding work} means the number of documents encoded with $f_t$.
Query encoding, ANN search, BM25, adapter fitting and application, and index builds are not
counted; they are orders of magnitude cheaper per call. Equal counts therefore do not mean
equal CPU, GPU or wall-clock totals, since strategies differ in these uncounted steps. Inline
encoding also adds query latency, which the count does not capture; Section~\ref{sec:res-cost}
measures it.

\textbf{Completion is a constraint.} The old index can be retired only once all $N$ documents
are encoded under $f_t$, so every admissible strategy must end there; one that does not get
there on its own is paired with a background re-encoder that completes. Total encoding work is
then $N$ for every admissible strategy. This follows from the constraint, not from a
measurement: any saving a reuse strategy appears to offer is a matter of \emph{when} the work is
paid, not whether. Section~\ref{sec:res-invariant} measures what does.

\subsection{When we measure quality, and what may win}
Quality is measured once per evaluation query, when that query is first served. A window
strategy encodes its query's missing window documents inline and is charged for them at that
moment; a rebuild serves the old index and is charged for whatever its background has done.
This rules out two accounting errors. Scoring window documents that are not yet encoded would
credit a window strategy with quality it has not paid for. Re-measuring one query at successive
checkpoints would reward a strategy for being measured late, after more background work.

For strategy $\sigma$ with mean served Recall@10 $\bar M_\sigma$ over the evaluation queries,
its \emph{debt} $\debt(\sigma) = M_T - \bar M_\sigma$ is the recall it forgoes while the
migration runs; lower is better. A rebuild serves source quality until it switches, so its debt
is the quality gap $M_T - M_S$ whenever the queries are served before then. On P2, for example,
the rebuild's background has encoded only 0.03 of $N$ when the evaluation queries are
served, so its debt is the full gap, 0.0471; window re-ranking has paid 0.264 of $N$
by then (Section~\ref{sec:model} prices it in advance) and its debt is +0.0023.
Section~\ref{sec:res-invariant} compares all six strategies.

\textbf{Which strategies may win.} A combination of strategy and background order is
\emph{admissible} on a pair if (i) its background order completes, (ii) it reaches target
quality, $\bar M_\sigma \ge M_T - \varepsilon$, and (iii) it serves no worse than the system it
replaces, $\bar M_\sigma \ge M_S$.

\textbf{Winner rule (pre-registered).} Among admissible combinations, those within $\tau$ of
the best debt are indistinguishable. Inside that band we order by cost-at-service, then by a
fixed simplicity ranking. We fixed $\varepsilon = 0.01$ and $\tau = 0.005$ before any pair was
run, and report the sensitivity of every winner to $\tau$ (Section~\ref{sec:res-select}).

\textbf{Two cost columns.} \emph{Total work} $C_{\mathrm{total}}$ is what a strategy ever
encodes; it is $N$ for everything admissible. \emph{Cost-at-service} $\Cserve$ is what it has
encoded by the time the evaluation queries are served. $\Cserve$ means different things in
different rows, so every row carries a basis label: for a window strategy it is work done; for a
rebuild that reached target quality without repairing anything, how far its background happened
to get; for a strategy that never reached target quality, $N$ by convention. Only rows with the
same basis are compared.

\subsection{Four quantities}\label{sec:quant}
\emph{Compatibility} says how close the two embedding spaces are: one minus the relative
residual of the semi-orthogonal Procrustes map $A$ (a rotation that allows the dimensions to
differ) carrying new-model document vectors onto old ones,
$\kappa = 1 - \lVert A(X_t) - X_s\rVert_F / \lVert X_s\rVert_F$, fitted on one document sample
and scored on a disjoint one. We also report linear CKA~\cite{cka}.

The \emph{window gap}, following~\cite{embedflow}, is the recall lost when the new model sees
only the old index's top-$K'$ instead of the whole corpus: $G(K') = M_T - M_{T\mid S_{K'}}$,
where $M_{T\mid S_{K'}}$ is the recall of $f_t$ re-ranking the window only. Both terms use exact
inner-product search, so $G$ carries no ANN error. Its counterpart without relevance judgments
is the \emph{residual miss}, the share of the new model's top-$K$ that falls outside the window:
$\rho(K') = 1 - \mathbb{E}_q\,\lvert\topk_K^t(q)\cap\topk_{K'}^s(q)\rvert/K$.

\emph{Coverage} is the cost side: the fraction of the corpus that appears in at least one
served window,
\[
\cov(K') = \tfrac{1}{N}\Bigl\lvert\bigcup_{q\in Q_{\text{served}}}\topk^s_{K'}(q)\Bigr\rvert .
\]
Every covered document must be encoded under $f_t$ by the time its query is served, so coverage
is what a window pays before the workload is served, less whatever the background reached
first. It needs the old index and nothing else. For $K' \le K$ the window \emph{is} the source
top-$K$, so $G(K) = M_T - M_S$ identically; window depths matter only above $K$.

\section{A Cost Model for Windowed Reuse}\label{sec:model}
\textbf{Scope.} \emph{Windowed reuse} means re-ranking under $f_t$ a candidate set nominated
without $f_t$'s document vectors, encoding the candidates it lacks first: S2--S5, not S0 or S1. We
write the model for S2, whose candidates are the source window $W_q = \topk^s_{K'}(q)$, so its
union is $\cov(K')$. Nothing below depends on how $W_q$ is nominated, so the same count over S3's
bridged-query windows or the unions of S4 and S5 bounds their inline work; for S4 and S5, $\cov(K')$ alone is
a lower bound.

A window migration's cost is $\cov(K')$, the union of the windows a workload touches: that is
the work done once \emph{every} query has been served. Pricing one before it runs also needs the
work done by any point in between, and needs to know when the obvious estimate is wrong. This
section gives both, from the serving index alone; Section~\ref{sec:plan} then decides with them.

\textbf{The multiplicity histogram.} A workload is a distribution $p$ over a set $\mathcal{Q}$ of distinct queries,
and arrivals $X_1, X_2, \dots$ are drawn i.i.d.\ from $p$ (our replays draw their streams this
way). Query $q$ has a source window $W_q$ of $K'$ documents. The \emph{window mass} of a document
is $P(d) = \sum_{q} p_q\,[d \in W_q]$, the probability that one arrival touches $d$. Let $U_m =
\bigcup_{i \le m} W_{X_i}$ and $\cov(m) = |U_m|/N$.

\begin{lemma}\label{lem:lower}
Any strategy that serves $X_1,\dots,X_m$ by re-ranking their source windows with $f_t$ has
encoded at least $|U_m|$ documents by then, and a strategy that encodes only on demand has
encoded exactly $|U_m|$.
\end{lemma}
\begin{proof}
Re-ranking $W_{X_i}$ with $f_t$ requires $f_t(d)$ for every $d \in W_{X_i}$, so every document of
$U_m$ has been encoded. On-demand encoding encodes nothing else.
\end{proof}

\begin{theorem}[Expected window-induced cost]\label{thm:curve}
$\mathbb{E}[\cov(m)] = 1 - \frac{1}{N}\sum_{d \in D} \bigl(1 - P(d)\bigr)^m.$
\end{theorem}
\begin{proof}
$d \notin U_m$ iff no arrival's window contains $d$. Each arrival misses $d$ independently with
probability $1 - P(d)$, so $\Pr[d \notin U_m] = (1-P(d))^m$. Sum over $d$ and divide by $N$.
\end{proof}

\begin{theorem}[The independent-window form is an upper bound]\label{thm:bound}
For every workload $p$ and every source index, $\sum_d P(d) = K'$, and
\[
\mathbb{E}[\cov(m)] \;\le\; 1 - \Bigl(1 - \tfrac{K'}{N}\Bigr)^{m},
\]
with equality for $m \ge 2$ if and only if $P(d) = K'/N$ for every $d$.
\end{theorem}
\begin{proof}
$\sum_d P(d) = \sum_q p_q\,|W_q| = K'$, so the mean window mass is $K'/N$ whatever $p$ and the
index are. The map $x \mapsto (1-x)^m$ is convex on $[0,1]$, and strictly convex for $m \ge 2$,
so by Jensen's inequality $\frac{1}{N}\sum_d (1-P(d))^m \ge (1 - K'/N)^m$, with equality iff $P$
is constant. Substitute into Theorem~\ref{thm:curve}.
\end{proof}

\begin{corollary}[Where the gap comes from]\label{cor:gap}
Expanding to second order around the mean mass, the excess of the bound over the expected
coverage is $\binom{m}{2}(1-K'/N)^{m-2}\,\mathrm{Var}_d[P(d)] + O(\mathbb{E}|P - K'/N|^3)$.
As $m \to \infty$ the expected coverage tends to the share of documents with $P(d) > 0$, the
union of the workload's windows, while the bound tends to 1.
\end{corollary}

The bound is exact only for an index that spreads windows perfectly evenly, and it is not a
modelling error to be corrected by a constant. Its excess is driven by the \emph{variance of
window mass}, which is large precisely when a few documents appear in many queries' windows: the
hubness of high-dimensional nearest-neighbour indexes~\cite{hubness}. This explains the
measurement of Section~\ref{sec:res-cost}, where the closed form overestimates at every depth and
for every encoder; Theorem~\ref{thm:bound} says it must.

\textbf{Why two results.} Theorem~\ref{thm:curve} is the price: exact in expectation, but it needs
the histogram, one retrieval pass over the servable queries. Theorem~\ref{thm:bound} needs only
$N$, $K'$ and $m$ and holds for any workload and index, so it is a guarantee that needs no
retrieval, and it says which way the obvious estimate errs: at $K'{=}50$ it gives 0.2946 of
$N$, against 0.2540--0.2653 measured (Section~\ref{sec:res-cost}).

\textbf{Procedure 1 (pricing a window migration).} Inputs: the source index $I_s$, the
servable queries $Q$, a window depth $K'$, and the arrival rate $\lambda$ per round. No
document is encoded with $f_t$.
(1)~Retrieve $\topk^s_{K'}(q)$ for every $q \in Q$ from $I_s$, and count for each document the
number $c(d)$ of windows that contain it. (2)~Form the \emph{multiplicity histogram}
$h_c = |\{d : c(d) = c\}|$; $\cov(K')$ is $1 - h_0/N$. (3)~Under a uniform workload,
$P(d) = c(d)/Q$, so Theorem~\ref{thm:curve} becomes
\[
\mathbb{E}[\cov(m)] = 1 - \tfrac{1}{N}\textstyle\sum_{c} h_c\,(1 - c/Q)^m ;
\]
under a skewed workload with known weights, accumulate $P(d)$ in step (1) instead.
(4)~The window-induced work after $r$ rounds is $N\,\mathbb{E}[\cov(\lambda r)]$, never more
than $N$ times the bound of Theorem~\ref{thm:bound}. Cost-at-service adds the background's
progress, which that bound does not cover: an order that repairs on demand encodes only inside
$U_m$ (Lemma~\ref{lem:lower}), while \emph{rand}, \emph{size} and \emph{signal} can encode
documents outside it. For those orders the total is a measurement, not a consequence of
Theorem~\ref{thm:bound}, and Section~\ref{sec:res-cost} reports it.
The histogram also explains why the background order matters only at low traffic intensity
(Section~\ref{sec:res-select}).

\begin{figure}[t]
\centering
\includegraphics[width=0.88\columnwidth]{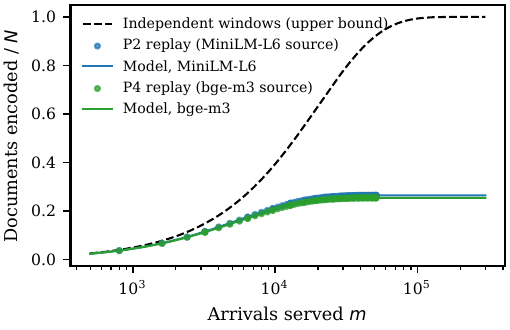}
\caption{Documents encoded by the on-demand window strategy against arrivals, two source encoders.
Dashed: the bound of Theorem~\ref{thm:bound}; solid: Theorem~\ref{thm:curve}, nothing fitted.}
\label{fig:curve}
\end{figure}

\textbf{Check against the replays.} The on-demand window strategy encodes exactly $U_m$
(Lemma~\ref{lem:lower}), so its trajectory is a realisation of $\cov(m)$. Fig.~\ref{fig:curve}
plots it for every round of every replay. All 850 points across the ten text pairs and
four traffic intensities lie below the bound of Theorem~\ref{thm:bound}, by 7--72\%
of its value. Theorem~\ref{thm:curve}, evaluated from each source encoder's histogram with
nothing fitted, predicts every point to within 0.81\%. The replays at $K'{=}25$
(Section~\ref{sec:res-plan}) add 387 points: 0 above the
bound, and the model within 0.74\%.

\section{The Migration Planner}\label{sec:plan}
Section~\ref{sec:model} prices a window at any depth from the serving index. What a
practitioner must still decide, since every strategy encodes $N$ documents either way, is
whether the upgrade is worth starting, what is served meanwhile, and when the work is paid. The planner estimates from statistics and never executes a candidate migration, as a
cost-based optimizer does. Its decisions restate conditions of the frame with the constants
declared in advance: D1 the band $\tau$; D2 the rule that a strategy serve no worse than the old
model, $\bar M_\sigma \ge M_S$; D3 the rule that it reach target quality,
$\bar M_\sigma \ge M_T-\varepsilon$, and the band. We call the version fixed before the replays the
\emph{committed planner}; ``the planner'' means it unless stated. It returns a family, a depth and a price. Coverage, which prices the migration, needs no target model
at all. The quality inputs do (Table~\ref{tab:planinputs}): $M_T - M_S$ on a judged sample of
documents, and $G(K')$ on the documents in the judged queries' source windows, at most $qK'$ for
$q$ judged queries ($2.5\%$ of $N{=}10^6$ for 500 queries at $K'{=}50$).

The planner cannot avoid running the new model on a sample, and that sample is not small
when a decision is close: a corpus
sample misstates the gap (Sections~\ref{sec:res-plan}, \ref{sec:scale}), and D1 decides as on the full corpus
for 10 of 12 pairs only at a 25\% sample (25.3\% of the
target encoding). But under completion they are not overhead either. Every document vector the
planner computes is one the migration needs, so when D1 says migrate it is the first instalment
of the bill, paid early; it is wasted only when D1 says no. Recomputed from 50 to 2,000 judged queries (5 seeds,
8 pairs), D1 decides as the full 3,490 do from 250 queries when the gap is
0.021 or more (5 pairs); D1 and D3 together are stable on 6 pairs
by 2,000, at most 9\% of $N$ in windows, and not by then on P2, P5, where the
gap or $G(25)$ sits at its threshold.

\begin{table}[t]
\centering
\caption{Planner inputs and what each costs.}
\label{tab:planinputs}
\small
\begin{tabular}{@{}lp{0.29\columnwidth}p{0.36\columnwidth}@{}}
\toprule
Input & Obtained from & Target encoding needed \\
\midrule
$\cov(K')$ & one retrieval pass over $I_s$ & \textbf{none}; 157--159\,s per encoder at $N{=}10^6$ \\
$M_S$ & retrieval over $I_s$ & none \\
$M_T$ & retrieval over a target-encoded sample & the sample (25\% of $N$ to match the full-corpus D1 on 10 of 12 pairs) and the queries \\
$G(K')$ & target re-ranking of the judged queries' source windows & the window documents, at most $q K'$ ($\le9\%$ of $N$ at $q{=}2,000$) \\
\midrule
\multicolumn{3}{@{}p{0.95\columnwidth}@{}}{Every document vector above is one a completed migration needs: a down payment when D1 says migrate, wasted only when it says no.}\\
\bottomrule
\end{tabular}
\end{table}

\begin{figure}[t]
\small
\hrule\smallskip
\textbf{Procedure 2 (the planner).} It uses Procedure~1 for the price. Inputs: $M_S$, $M_T$ and each adapter's $M_A$ on a judged
sample; $G(K')$ for $K' \in \{10, 25, 50, \dots\}$; $\cov(K')$ from $I_s$. Output: a family, a
depth and a price.
\begin{enumerate}\setlength{\itemsep}{0pt}
\item[D1] $g \leftarrow M_T - M_S$. If $g \le \tau$: no reuse; rebuild.
\item[D2a] A bridge is a candidate only if $M_A - M_S > \tau$.
\item[D3] $K' \leftarrow$ the smallest depth $> K$ with $G(K') \le \varepsilon$ and $g - G(K') > \tau$.
If none: rebuild.
\item[D2b] If $1 - \cov(K') \le 0.01$ (a cut-off for ``nothing to defer'', not $\varepsilon$): rebuild. Otherwise window S2 at
$K'$, paying $\cov(K')\,N$ by the time the workload is served.
\end{enumerate}
As committed, S4 and S5 are never candidates.\smallskip\hrule
\end{figure}

Procedure~2 formalizes the three decisions and orders them by what each depends on. D1 comes first, because nothing else
matters if the upgrade is not worth making. D2 is split: whether a bridge is a candidate (D2a)
needs only $M_A$, but whether a window has anything to defer (D2b) needs the depth D3 picks. On the running example P2, $g = 0.0471> \tau$, so D1 migrates; the best
bridge of any family misses the old model by $-0.27$, so D2 admits none; $K'{=}10$ is
inadmissible ($G = 0.0471$, the whole gap) and $K'{=}25$ is admitted ($G = 0.0094$);
the plan is S2 at $K'{=}25$, paying 0.149 of $N$ (Table~\ref{tab:plan}).

\textbf{D1: worth more than a rebuild?} The upgrade is worth disrupting only if the new model is better by more than the margin one
has declared equivalent. Under first-arrival measurement this is not a
proxy for anything. A rebuild serves source quality throughout, so its debt \emph{is} the
quality gap, and it equals $M_T-M_S$ exactly on 11 of our 13 pairs. D1
therefore does not stop a complete migration: when $g \le \tau$ the rebuild is within the band,
so no reuse strategy is worth its cost, and skipping the upgrade loses nothing distinguishable.

\textbf{D2: which family?} The bridge clause follows from the frame: a
bridge serves $M_A$ until the background finishes, so it beats not migrating only if $M_A >
M_S$, and distinguishably so only by more than $\tau$. In our data it never
admits a bridge: the one pair whose bridges clear $M_S$, P5r, clears it by less than $\tau$
(Section~\ref{sec:res-map}), so the released planner lists S1 and S3 as excluded; P5r's replay
shows that this can exclude a winning S3. The second clause is a scope condition, and we
state it as one:

\begin{quote}\itshape
A cost-based planner recommends only strategies whose cost it can bound from its inputs.
\end{quote}

The committed planner computed $\cov(K')$ only over S2's windows, which for S4 and S5 is
a lower bound (Section~\ref{sec:model}). We exclude them because the planner as committed cannot
price them, not because they never win; Section~\ref{sec:res-plan} lists every cell they win.

\textbf{D3: at what depth?} The cost of the chosen depth is $\cov(K')$ once the workload is served, and
Section~\ref{sec:model} gives it at any point before that. Both conditions are needed. A window serves target
quality minus its gap, so a shallower window is cheaper precisely because it is worse.
Admissibility alone would accept a depth whose gap consumes the whole value of the upgrade;
the second condition, $g - G(K') > \tau$, requires the window to keep more than $\tau$ of the
upgrade's gain $g$.

\section{Experimental Setup}\label{sec:setup}
\textbf{Operating point, and why it can falsify the claims.} At the primary traffic
intensity the rebuild's background has encoded only 0.03 of the corpus by the time
the evaluated population is served, so a window strategy that looks cheap must be cheap in its
own inline encoding, not because the background did the work. On text, windows at $K'{=}50$ touch only
0.2540--0.2653 of the corpus, so there is room both for a deferral to show and for
the cost model to be wrong. The code corpus is the opposite case: its windows touch every
function from $K'{=}50$, so window selection is vacuous by construction, and we use it as a
boundary test (Section~\ref{sec:res-regime}), not as evidence about selection.

\textbf{Text.} MS MARCO passages~\cite{msmarco}, a $10^6$-passage subsample built from
\emph{every} judged passage (MS MARCO dev, TREC-DL 2019/2020~\cite{trecdl19,trecdl20}) plus a uniform random fill. A
random subsample would drop most judged passages and make $M_T$ and $G$ meaningless. Of the
6,980 judged dev queries, a seeded half (3,490) is held out for evaluation. The rest form
traffic only. Evaluation queries are not held out of the stream: they arrive like traffic, and
each is scored once, at its first service, before its own repair can warm its window. The 97 TREC-DL
queries, which are deeply judged, are evaluation-only.
For the scale study (Section~\ref{sec:scale}) we encode the full collection, all
8,841,823 passages, with the models of P2, P3 and P9 (P2f, P3f, P9f), and measure the planner's
static inputs over the same queries, then replay the three pairs at 2M, 4M and the full size.

\textbf{Code.} CodeSearchNet Python~\cite{codesearchnet}, 45,773 functions with 43,730
servable queries and 3,000 evaluation queries.

\textbf{Workload and traffic.} Arrivals are uniform over the servable queries, and every
workload-dependent claim below is scoped to uniform arrivals. Uniformity fixes the coverage
\emph{numbers} we report; it does not enter the identity between the on-demand strategy's cost
and the coverage of the served windows, nor Theorem~\ref{thm:bound}, which holds for every
workload. With background budget $b$ encodings per round and
$\lambda$ queries per round, traffic intensity is $\mathrm{tr} = \lambda K'/b \in
\{4,16,64,256\}$; $\mathrm{tr}{=}64$ is pre-registered as primary.

\textbf{Skew, a sensitivity check.} A skewed workload changes the window masses that Procedure~1
reads, not the procedure. Replayed under a pre-registered Zipf(0.9) workload, P2, P5, P11, P13 keep all
2,088 points below Theorem~\ref{thm:bound}'s bound and S2's cost within 0.00005 of
$N$; but quality is workload-weighted, so debts move by up to 0.0070 and the winner
changes on P2 (to S4) and P5 (to S0); weighted by Zipf, P5's gap is negative. We then replayed 36 more
pre-registered Zipf workloads ($\alpha = 0.3$, 0.6, 1.2 on the ten text pairs; 0.9 on the other
six): the identity holds on all 36, and the committed family matches on 32, missing P5 at three
skews and P14 at $\alpha = 1.2$.{} A static, post hoc scan of
all 10 text pairs (five popularity orders per $\alpha$) shows why: the weights shrink the 3,490
judged queries to an effective 2,896 at $\alpha{=}0.3$, 796 at 0.6,
93 at 0.9 and 18 at 1.2, and D1/D3 then decide otherwise than under uniform arrivals in
3, 8, 22 and 27 of 50 settings
(P2: 13 of 20). It reproduces each replayed rebuild debt (difference at most $10^{-15}$).

\textbf{Pairs.} Table~\ref{tab:pairs} lists the fifteen pairs of the main replay grid; the sixteenth, P5r
(Qwen3-Embedding-0.6B$\to$8B), tests same-family LLM bridges (Table~\ref{tab:adapters}) and was
replayed after the others, outside their counts (Section~\ref{sec:res-map}). P0 is a control: the same model with
its approximate index rebuilt under another seed. The ten text pairs span five BERT-lineage
encoders (384--1024 dimensions): MiniLM-L6~\cite{minilm,sbert}, bge-small and bge-base~\cite{bge},
bge-m3~\cite{bgem3} and e5-base~\cite{e5}. P7 and P8, on the code corpus with CodeSage~\cite{codesage}, are negative controls for
deferral, not evidence about family selection (Section~\ref{sec:res-regime}). P5p and P5q pair a
BERT-lineage source with Qwen3-Embedding-8B~\cite{qwen3emb} (4096 dimensions) on the same 1M text corpus.
Both are replayed at $K'{=}50$ like the text pairs: the window wins on P5q; on
P5p, whose $G(50)$ exceeds $\varepsilon$, the lexical union does. D3 would put P5q at
$K'{=}25$ and P5p at $K'{=}100$, deeper than the replay, so P5p's plan is untested.{} Under the final
encodings two pairs are not upgrades: P14's gap is $-0.0009$ and P8's is $+0.0008$.
Rather than drop them, we keep them as the cases D1 exists to catch.

\begin{table*}[t]
\centering
\caption{Model pairs. Recall@10 on held-out queries; $\rho$ and $G$ at $K'{=}50$; d$\to$r:
demand-then-rand. $^\ddagger$Procrustes residual above 1: a compatibility failure.}
\label{tab:pairs}
\footnotesize\setlength{\tabcolsep}{3.2pt}
\begin{tabular}{@{}llllllllllll@{}}
\toprule
ID & source $\to$ target & type & $M_S$ & $M_T$ & gap & $M_A$ & $\kappa$ & CKA & $\rho(50)$ & $G(50)$ & replay winner \\
\midrule
P0 & bge-base$\to$bge-base & control & $0.8805$ & $0.8805$ & $+0.0000$ & $0.8805$ & $+1.000$ & $1.000$ & -- & -- & S0/rand \\
P2 & MiniLM-L6$\to$bge-base & cross-family & $0.8335$ & $0.8805$ & $+0.0471$ & $0.4119$ & $+0.037$ & $0.617$ & $0.215$ & $+0.0022$ & S2/d$\to$r \\
P3 & bge-small$\to$bge-base & same family & $0.8652$ & $0.8805$ & $+0.0154$ & $0.6832$ & $+0.391$ & $0.770$ & $0.140$ & $-0.0029$ & S2/d$\to$r \\
P4 & bge-m3$\to$bge-base & same family & $0.8586$ & $0.8805$ & $+0.0219$ & $0.6069$ & $+0.289$ & $0.721$ & $0.226$ & $-0.0050$ & S2/d$\to$r \\
P5 & bge-m3$\to$bge-small & same family & $0.8586$ & $0.8652$ & $+0.0066$ & $0.4564$ & $+0.276$ & $0.713$ & $0.245$ & $-0.0069$ & S2/d$\to$r \\
P9 & MiniLM-L6$\to$bge-small & cross-family & $0.8335$ & $0.8652$ & $+0.0317$ & $0.4958$ & $+0.044$ & $0.653$ & $0.192$ & $-0.0010$ & S2/d$\to$r \\
P10 & MiniLM-L6$\to$bge-m3 & cross-family & $0.8335$ & $0.8586$ & $+0.0251$ & $0.3252$ & $+0.025$ & $0.585$ & $0.292$ & $-0.0077$ & S2/d$\to$r \\
P11 & MiniLM-L6$\to$e5-base & cross-family & $0.8335$ & $0.8796$ & $+0.0462$ & $0.0491$ & $-0.090$$^\ddagger$ & $0.608$ & $0.211$ & $+0.0037$ & S4/d$\to$r \\
P12 & bge-m3$\to$e5-base & cross-family & $0.8586$ & $0.8796$ & $+0.0210$ & $0.3509$ & $+0.308$ & $0.734$ & $0.222$ & $-0.0019$ & S2/d$\to$r \\
P13 & bge-small$\to$e5-base & cross-family & $0.8652$ & $0.8796$ & $+0.0144$ & $0.4510$ & $+0.400$ & $0.750$ & $0.168$ & $+0.0008$ & S2/d$\to$r \\
P14 & bge-base$\to$e5-base & cross-family & $0.8805$ & $0.8796$ & $-0.0009$ & $0.7632$ & $+0.442$ & $0.817$ & $0.113$ & $-0.0018$ & S0/rand \\
\midrule
P7 & MiniLM-L6$\to$codesage-s & general$\to$code & $0.9590$ & $0.9897$ & $+0.0307$ & $0.8559$ & $+0.064$ & $0.390$ & $0.372$ & $+0.0114$ & S2/rand \\
P8 & codesage-s$\to$codesage-b & code$\to$code & $0.9894$ & $0.9902$ & $+0.0008$ & $0.9701$ & $-0.029$ & $0.539$ & $0.145$ & $-0.0001$ & S0/rand \\
\midrule
P5p & MiniLM-L6$\to$Qwen3-Emb-8B & enc.$\to$LLM & $0.8335$ & $0.9069$ & $+0.0735$ & $0.5721$ & $+0.155$ & $0.591$ & $0.248$ & $+0.0118$ & S4/d$\to$r \\
P5q & bge-base$\to$Qwen3-Emb-8B & enc.$\to$LLM & $0.8805$ & $0.9069$ & $+0.0264$ & $0.7789$ & $+0.303$ & $0.719$ & $0.183$ & $+0.0027$ & S2/d$\to$r \\
\bottomrule
\end{tabular}
\end{table*}

\textbf{Encoders.} Every model is encoded once under an explicit contract (prompt, pooling,
padding side, EOS, revision) and cached, so all strategies replay over identical vectors. The
checks we can make pass for every encoder: padding invariance (cosine 1.0), self-retrieval
1.0, and 23 structural checks per model with no failure.

\textbf{Hardware.} One RTX 4090 D (48 GB), 20 vCPUs (Xeon Platinum 8481C), 90 GB RAM. Counts, coverage,
recall and debt do not depend on the hardware; latency and GPU-hours do.

\textbf{Pre-registration and provenance.} The winner rule, $\varepsilon$, $\tau$, the traffic
grid and the primary setting were committed before any pair in the reported round ran. The ten
text pairs share one configuration hash. The measurement code is identical across the three commits that produced the replays. Window
measurements use exact inner-product search; HNSW~\cite{hnsw} is used only for P0. Every replay passes sanity invariants that hold by construction (e.g., $G{=}0$ when the window
is the corpus; identical arrival sequences across strategies).

\section{Results}\label{sec:results}
Table~\ref{tab:claims} maps each finding to its anchor. Each replay is a deterministic run and the
tables report seed 0. Every windowed pair was run at 5 seeds, seed 0 and
4 more (9 windowed pairs $\times$ 5 seeds $=$ 45 replays). The winner among the families D2 admits matched the plan in
45 of 45; S2's background orders tied within $\tau$ in 45 of 45; S2's debt moved by
at most 0.0057 and its cost by 0.00005 of $N$. The unpriced lexical union is
seed-sensitive: counting seeds, it is the raw winner on P2 (1 of 5), P9 (1 of 5), P11 (4 of 5) and beats S2's debt by more
than $\tau$ on P2 (1 of 5), P9 (1 of 5), P11 (5 of 5).

\subsection{What separates strategies once total encoding work cannot}\label{sec:res-invariant}
A finished migration has encoded the whole corpus, so total encoding work is $N$ for every admissible
strategy by definition. The identity is trivial once stated; what the cited work omits is
the price of the work left after a deferral. All 180 replayed strategy--order combinations that use the
\emph{demand} order (every demand cell of the replay grid) never finish and are inadmissible by definition. Every saving a reuse strategy offers is therefore a
deferral, and what decides whether the deferral is worth having is the spread in
Fig.~\ref{fig:space}. Window strategies
pay roughly a quarter of the corpus before the workload is served at target quality, and their
debt stays within $\pm0.0057$ of the target. Lexical unions pay more than three times
as much; at seed 0 their quality clears the equivalence band on one pair, P11, by 0.000086
(Section~\ref{sec:res-plan}); across seeds, on 3 pairs at least once. The bridges never reach target quality.

\subsection{Bridges must beat the old model, not approach the new one}\label{sec:res-map}
Adapter bridges are evaluated in their literature by recall \emph{relative to the new model}:
Drift-Adapter reports 95--99\% of the Recall@10 of full re-embedding~\cite{drift-adapter}. For
the upgrade decision the relevant comparison is with \emph{not migrating}. A bridge serves $M_A$
while the background runs, so it is worth deploying only if $M_A > M_S$, that is, only if its
relative recall $M_A/M_T$ exceeds $M_S/M_T$. On our 12 upgrades that break-even point, $M_S/M_T$, is
91.9--99.2\%. A bridge that recovers 95--99\% of the new model can therefore still be worse
than leaving the old index alone, and the better the old model, the more likely this is.

\begin{figure}[t]
\centering
\includegraphics[width=0.88\columnwidth]{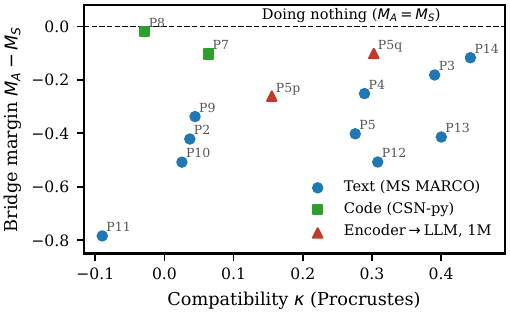}
\caption{Procrustes-bridged minus source quality against compatibility: every pair is below zero.
P5r is in Table~\ref{tab:adapters}.}
\label{fig:mapping}
\end{figure}

\textbf{The bridge we replay.} The bridge in our replays is our reimplementation of a
semi-orthogonal Procrustes map, one of the three families Drift-Adapter evaluates, fitted on
10,000 random document pairs, with no query or judgment (Table~\ref{tab:adapters} reports
each family's best over three training sets, so its entries can be higher). Its relative recall
spans 6\% (P11) to 98\% (P8), median
60\%. On P8 it reaches 98.0\% (Table~\ref{tab:adapters}'s 99 is the best over three
training sets), inside Drift-Adapter's reported range, so the
reimplementation reproduces the published yardstick; yet
the break-even point there is 99.9\%, so the bridge serves below the old model. Across all
14 non-control pairs of the main grid, including the encoder$\to$LLM pairs P5p and P5q, it serves between
0.0192 (P8) and 0.7843 (P11) Recall@10 \emph{below the
source model} (Fig.~\ref{fig:mapping}); the best family of Table~\ref{tab:adapters} still falls short
on P8, by 0.012. Over the full $\tau$ sweep, S1 and S3 are selected in
0 of 416 cells.

\textbf{Every family falls short of the band.} Table~\ref{tab:adapters} reports the families
Drift-Adapter evaluates (Procrustes, low-rank affine, a residual MLP) plus ridge affine with a
held-out penalty, each trained on document pairs, on traffic query pairs and on both, never on
evaluation queries. Ridge affine is the strongest family on 6 of the 10 text
pairs and Procrustes on 3; the MLP, which starts from the affine fit, stays within
0.025 of it. On P14 the MLP reaches 95\% relative recall, at the
edge of Drift-Adapter's reported range, yet its margin over the old model is $-0.045$.
Of the 15 pairs in Table~\ref{tab:adapters} (every pair of the main grid but the control P0,
plus P5r), one has bridges that beat the old model: P5r, Qwen3-Embedding-0.6B$\to$8B,
two LLM embedders of one family, where Procrustes, affine and the MLP clear $M_S$ by
+0.0027--+0.0044 at 96\% relative recall. That recovers 11\% of the pair's
gap and falls short of $\tau$ by 0.0006, so D2 admits no bridge on any pair (a replay
serves $M_A$ until the switch, so it cannot change that); elsewhere the
best margin is $-0.012$ (P8, code). Replayed afterwards, P5r confirms the bridge's
verdict (replayed Procrustes: $M_A - M_S = +0.0016$; best static family +0.0044; both below
$\tau$) and exposes a limit of D2: the mapped window's debt, $0.0050$ at $0.255$ of $N$, is within
$\tau$ of S2's $0.0016$ at $0.266$, so S3 wins on cost. The point is the bar, not Drift-Adapter's
own benchmarks: it is $M_S$, and relative recall cannot show that a bridge clears it.

\begin{table}[t]
\centering
\caption{Adapter families: best relative recall $M_A/M_T$ (\%) over three training sets, and best
$M_A-M_S$.}
\label{tab:adapters}
\footnotesize\setlength{\tabcolsep}{3pt}
\begin{tabular}{@{}lccccccc@{}}
\toprule
Pair & $M_S$ & Procr. & affine & l-r 64 & l-r 128 & MLP & best $M_A{-}M_S$ \\
\midrule
P2 & $0.833$ & $47$ & $64$ & $25$ & $50$ & $63$ & $-0.272$ \\
P3 & $0.865$ & $78$ & $73$ & $37$ & $61$ & $72$ & $-0.180$ \\
P4 & $0.859$ & $71$ & $62$ & $21$ & $46$ & $60$ & $-0.234$ \\
P5 & $0.859$ & $55$ & $47$ & $18$ & $37$ & $48$ & $-0.383$ \\
P5p & $0.833$ & $63$ & $74$ & $30$ & $59$ & $73$ & $-0.165$ \\
P5q & $0.881$ & $86$ & $86$ & $36$ & $70$ & $85$ & $-0.101$ \\
P5r & $0.865$ & $96$ & $96$ & $60$ & $86$ & $96$ & $+0.004$ \\
P7 & $0.959$ & $89$ & $84$ & $63$ & $79$ & $86$ & $-0.075$ \\
P8 & $0.989$ & $99$ & $99$ & $84$ & $96$ & $98$ & $-0.012$ \\
P9 & $0.833$ & $57$ & $78$ & $38$ & $67$ & $79$ & $-0.151$ \\
P10 & $0.833$ & $37$ & $50$ & $18$ & $38$ & $47$ & $-0.404$ \\
P11 & $0.833$ & $6$ & $64$ & $25$ & $50$ & $63$ & $-0.270$ \\
P12 & $0.859$ & $46$ & $65$ & $21$ & $47$ & $65$ & $-0.287$ \\
P13 & $0.865$ & $55$ & $72$ & $36$ & $59$ & $69$ & $-0.234$ \\
P14 & $0.881$ & $89$ & $95$ & $44$ & $82$ & $95$ & $-0.045$ \\
\bottomrule
\end{tabular}
\end{table}

\textbf{Compatibility predicts how badly bridging fails, not whether it does.} Within a corpus a
cheap compatibility score tracks bridge quality: over the ten text pairs $r(\kappa, M_A) =
+0.755$ (CKA: $+0.754$; $\kappa$ with the two LLM-target pairs: $+0.712$). Neither
measure transfers across corpora: pooling the code pairs takes $r(\mathrm{CKA}, M_A)$ to
$-0.186$. The bridge misses target quality, $M_A \ge M_T - \varepsilon$, on every pair:
$M_A-(M_T-\varepsilon)$ runs from $-0.0100$ (P8) to $-0.8205$
(P11), and is $-0.1065$ even at the highest $\kappa$ (0.4424,
P14). (Table~\ref{tab:adapters}'s last column measures against $M_S$ instead.) Our data lie on one side of any success
boundary, $\kappa \in [-0.090, 0.442]$, so we fit none.  Compatibility cannot replace measuring $M_A$ against
$M_S$.

\subsection{The cost of windowed reuse is known in advance}\label{sec:res-cost}
\textbf{The bound.} To serve a query by re-ranking its source window, every document in that
window must be encoded under $f_t$. A window strategy serving $Q_{\text{served}}$ must
therefore encode at least $N\cdot\cov(K')$ documents before it serves that population at
target quality, and can defer at most $1-\cov(K')$ of the work. It is a counting argument, computable before any encoding.

\textbf{The replay attains the bound.} Table~\ref{tab:econ} compares the work each window strategy had
actually done when it served the evaluated population with the static $\cov(50)$ of its source
encoder. On all 10 pairs, across 4 source encoders, the replay stays
below the bound by 31--50 documents in $10^6$, and never exceeds it. The
residual is the handful of windows whose traffic queries had not yet arrived. Since a window
strategy encodes the union of the windows it serves by construction, this confirms the
implementation rather than a prediction. What it adds is that background work does not push
the realised cost above the bound.

\begin{table}[t]
\centering
\caption{Window cost-at-service (replay, primary setting) against the static $\cov(50)$ of its source
encoder.}
\label{tab:econ}
\footnotesize\setlength{\tabcolsep}{3.5pt}
\begin{tabular}{@{}lp{0.19\columnwidth}cccc@{}}
\toprule
Source & Pairs & $\Cserve/N$ & $\cov(50)$ & gap (docs) & rel. \\
\midrule
MiniLM-L6 & P2, P9, P10, P11 & $0.264224$ & $0.264274$ & 50 & $0.019\%$ \\
bge-base & P14 & $0.263661$ & $0.263695$ & 34 & $0.013\%$ \\
bge-m3 & P4, P5, P12 & $0.253968$ & $0.254009$ & 41 & $0.016\%$ \\
bge-small & P3, P13 & $0.261183$ & $0.261214$ & 31 & $0.012\%$ \\
\bottomrule
\end{tabular}
\end{table}

\begin{figure*}[t]
\centering
\includegraphics[width=0.88\textwidth]{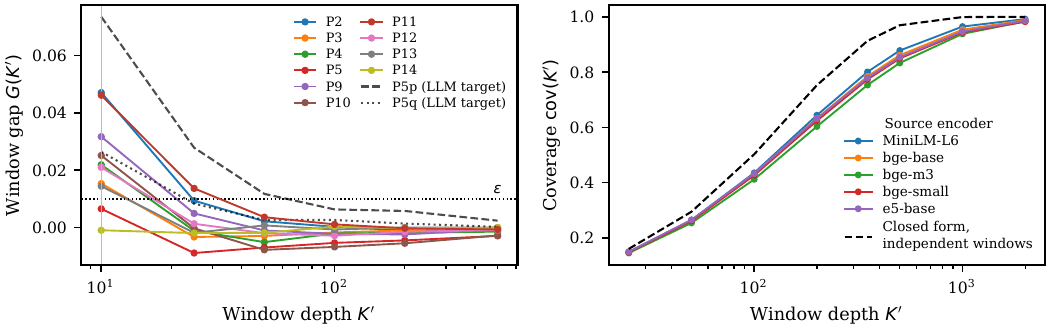}
\caption{(a) Window gap $G(K')$: ten text pairs (solid), two encoder$\to$LLM pairs (dashed, dotted).
(b) Coverage against depth, five source encoders, and the independent-window form.}
\label{fig:frontier}
\end{figure*}

\textbf{The independent-window form overestimates.} If queries drew windows independently,
coverage would be $1-(1-K'/N)^{|Q|}$, computable from three integers. That baseline
\emph{overestimates} cost at every depth and for every encoder, as Theorem~\ref{thm:bound}
requires (Fig.~\ref{fig:frontier}b),
because real windows overlap more than independent draws do. At $K'{=}50$ it predicts
encoding 0.2946 of the corpus against a measured 0.2540--0.2653, overstating
cost by 11.1--16.0\%. A planner using the closed form would under-provision the deferrable work by
that margin, most at $K'{=}200$ (16.9--24.8\%). Coverage barely depends on
the encoder (0.2540--0.2653 at $K'{=}50$): it prices a migration but cannot tell pairs apart.

\textbf{The deferral is paid in latency.} S0 adds nothing to a query and S1 one forward pass plus
a matrix product; S2--S5 also encode, in one batched pass before answering, the query's window
documents not yet encoded: on average 36.4--37.8 at $K'{=}50$ and
21.3 at $K'{=}25$, fewer as the migration proceeds. On one
RTX 4090 D, that takes 31--187\,ms for 50 documents and
16--93\,ms for 25, plus at most 119--665\,ms behind a
background batch.{} A deployment that cannot afford that wait should rebuild or take the
shallower window D3 allows.

\subsection{How deep a window?}\label{sec:res-depth}
Fig.~\ref{fig:frontier}a gives the frontier ($G(50)$ per pair in Table~\ref{tab:pairs}, $G(25)$ in Table~\ref{tab:d3}). At $K'{=}25$ the gap
exceeds $\varepsilon$ on 1 text pair (P11, $G = +0.0137$);
by $K'{=}50$ the largest gap is $+0.0037$ (P11), and from $K'{=}100$ on it never
exceeds $+0.0011$.

The pre-registered \emph{global} rule takes the smallest depth admissible on every pair,
$K^\ast{=}50$, which every text replay used; P11 alone sets it, so it is conservative rather than representative. D3 instead
chooses \emph{per pair}: $K'{=}25$ for 8 of the nine pairs that migrate,
and $50$ for P11. Coverage roughly halves between the two depths, across the five encoders' indexes, from 0.2540--0.2653
to 0.1452--0.1499 (the upper end is e5-base's index, whose $\cov(25)$ is 0.1499 against bge-base's
0.1490; no pair in Table~\ref{tab:d3} has an e5-base source). A per-pair planner therefore promises to pay roughly half as much before the workload is served
at target quality; Section~\ref{sec:res-plan} tests that.

\textbf{Negative gaps are most likely a judgment artefact,} not a real advantage of the window:
the deeply judged TREC-DL queries move $G$ toward zero. On 7 of the ten text pairs $G(50)$
is negative: the window, restricted to source candidates, scores \emph{above} exhaustive target
retrieval. Sparse judgments explain that more simply than a cascade effect: MS MARCO dev has about one
judged passage per query, the window keeps retrieval near the judged pool, and exhaustive target
retrieval surfaces unjudged documents that count as misses. The deeply judged TREC-DL queries
separate the two readings (Table~\ref{tab:sparsity}). Where dev reports $G$ down to
$-0.0029$, TREC-DL reports $-0.0004$ to $+0.0014$, close to zero: the window matches the target model rather than beating it. The evidence is
limited: a sign check on 97 queries at $K'{=}500$, and at the planner's depths only P11
and P14 were re-measured, with mixed results (on P11, TREC-DL puts $G(50)$ at
$+0.0136$, above $\varepsilon$). Sparse judgments may understate a positive gap, D3's
exposure.

\begin{table}[htbp]
\centering
\caption{$G$ at $K'{=}500$: sparse dev judgments against deeply judged TREC-DL (97
queries); pairs with both zero omitted.}
\label{tab:sparsity}
\small
\begin{tabular}{@{}lrr@{}}
\toprule
Pair & $G$ on dev (Recall@10) & $G$ on TREC-DL (nDCG@10) \\
\midrule
P5 & $-0.0029$ & $-0.0004$ \\
P10 & $-0.0027$ & $-0.0000$ \\
P4 & $-0.0014$ & $+0.0000$ \\
P12 & $-0.0009$ & $-0.0003$ \\
P2 & $-0.0007$ & $+0.0001$ \\
P11 & $-0.0006$ & $+0.0014$ \\
P9 & $-0.0006$ & $+0.0008$ \\
\bottomrule
\end{tabular}
\end{table}

\subsection{Does the planner decide correctly?}\label{sec:res-plan}
\textbf{The family decision is consistent with the replay on 12 of 13
pairs}, 9 of the 10 text pairs; those replays predate the planner, so this
is consistency, not prediction. The deciding function receives the
static quantities and nothing else
(Table~\ref{tab:plan}). P0 is the identity control, and on the code
pairs coverage leaves no deferral to price, so those three carry little information about
selection. D1 sends the three pairs whose gap sits inside the band ($-0.0009$ (P14), $+0.0000$ (P0), $+0.0008$ (P8)) to a
rebuild, and the replay agrees; any threshold between $+0.0008$ and $+0.0066$
(P5) classifies every pair, and the $\tau$ fixed in advance falls there. Only two
of the three are non-trivial negatives, so we do not offer them as evidence for D1, whose
justification is the identity $\debt(\text{S0}) = M_T - M_S$. A rule that always answers S2
agrees on 9 of 13; the planner's further agreements are exactly the
rebuild pairs. Scored at the other traffic intensities, the family holds on 13 of
13 at $\mathrm{tr}{=}16$ and 13 of 13 at 256, and on
10 of 13 at 4 (misses: P0, P14, P7), where background progress decides.

The exception, P11, is the pair on which the replay selects S4, which D2 excludes by
rule. There S4 pays 0.8568 of $N$ against the window's 0.2642
(3.24$\times$) for a debt advantage of 0.005086. That clears the equivalence band
by 0.000086, or 1.7\% of its width, and widening $\tau$ by that much
selects S2 (Fig.~\ref{fig:tau}). At seed 0 the band decides by 0.000086; across 5 seeds S4's advantage is
0.0051--0.0099 and clears $\tau$ on every seed: not noise.

\begin{table*}[t]
\centering
\caption{Planner decisions from static inputs against the replay. Planner cost: $\cov(K')$ at D3's
depth; replay cost: $\Cserve/N$ at the replay's depth. $^\ast$Priced on the code corpus.}
\label{tab:plan}
\footnotesize\setlength{\tabcolsep}{4pt}
\begin{tabular}{@{}lrllrrlrrl@{}}
\toprule
& & \multicolumn{4}{c}{planner (static)} & \multicolumn{3}{c}{replay} & \\
\cmidrule(lr){3-6}\cmidrule(lr){7-9}
Pair & gap & D1 & D2 & D3 $K'$ & cost & selected & $K'$ & cost & agree \\
\midrule
P0 & $+0.0000$ & rebuild & S0 & -- & -- & S0/rand & 50 & $0.030$ & \ding{51} \\
P2 & $+0.0471$ & migrate & S2 & 25 & $0.149$ & S2/d$\to$r & 50 & $0.264$ & \ding{51} \\
P3 & $+0.0154$ & migrate & S2 & 25 & $0.148$ & S2/d$\to$r & 50 & $0.261$ & \ding{51} \\
P4 & $+0.0219$ & migrate & S2 & 25 & $0.145$ & S2/d$\to$r & 50 & $0.254$ & \ding{51} \\
P5 & $+0.0066$ & migrate & S2 & 25 & $0.145$ & S2/d$\to$r & 50 & $0.254$ & \ding{51} \\
P9 & $+0.0317$ & migrate & S2 & 25 & $0.149$ & S2/d$\to$r & 50 & $0.264$ & \ding{51} \\
P10 & $+0.0251$ & migrate & S2 & 25 & $0.149$ & S2/d$\to$r & 50 & $0.264$ & \ding{51} \\
P11 & $+0.0462$ & migrate & S2 & 50 & $0.264$ & S4/d$\to$r & 50 & $0.857$ & excl.\ (S4) \\
P12 & $+0.0210$ & migrate & S2 & 25 & $0.145$ & S2/d$\to$r & 50 & $0.254$ & \ding{51} \\
P13 & $+0.0144$ & migrate & S2 & 25 & $0.148$ & S2/d$\to$r & 50 & $0.261$ & \ding{51} \\
P14 & $-0.0009$ & rebuild & S0 & -- & -- & S0/rand & 50 & $0.030$ & \ding{51} \\
P7 & $+0.0307$ & migrate & S2 & 100 & $1.000$$^\ast$ & S2/rand & 200 & $1.000$ & \ding{51} \\
P8 & $+0.0008$ & rebuild & S0 & -- & -- & S0/rand & 200 & $1.000$ & \ding{51} \\
\bottomrule
\end{tabular}
\end{table*}

\textbf{The priced decision, tested before its replays.} The plan was committed at
15:16 (instance clock) and the replays at D3's depth started after it; the run
log and the planner check record the order. Table~\ref{tab:d3} scores the 8 pairs D3 sent
to $K'{=}25$ on two registered criteria, price (the window's cost-at-service attains $\cov(25)$ to
within 0.1\%) and quality (its debt lies within $\tau$ of $G(25)$), and we also report the family
(does S2 win?). Price holds on every pair where it can be scored (7 of 8),
quality on 7, family on 6.

\emph{The price holds.} On every pair where the window reaches target quality (7 of
8), it has encoded $\cov(25)$ of $N$, to within 0.00004, when the evaluated
population is served, against 0.254--0.264 at $K'{=}50$; inline repairs fall to
21.3 per evaluated query (p95: 25). On P2 the
window serves below $M_T - \varepsilon$, so by the frame's definition it never reaches target
quality and has no cost-at-service to score.

\emph{The quality holds on 7 of 8.} P5 misses the tolerance by
0.00006. Over all 18 replays at both depths, the replay's debt is never below
the static $G(K')$ and exceeds it by up to 0.0058, most on the two pairs whose static
gaps are most negative (P10 at 50 and P5 at 25). We have not isolated the cause; one candidate is that a served answer also
admits the best documents already encoded, which a static window re-rank does not. So on these MS MARCO pairs $G$ is a
lower bound on the replay's debt, with a measured margin, and $\varepsilon = 0.01$ does not leave
that margin: on P2 it admits $K'{=}25$, whose debt never reaches target quality. Re-running D3
with $\varepsilon - 0.006= 0.004$ moves P2 (to 50) and P9 (to 50); every window it then admits
reaches target quality (9 of 9) and S2 wins on all but P11. This margin is calibrated in-sample, on the MS MARCO replays; on NQ, out of sample, debt $-\,G$ is two-sided,
$[-0.0060, +0.0050]$, and only its upward side, inside the margin, can push debt past $\varepsilon$.

\emph{The family holds on 6 of 8.} On P2 and P9 the shallow
window's debt (0.0102 and 0.0063) exceeds $\tau$, and the lexical union,
which D2 does not admit, buys it back: S4 wins by 0.0073--0.0105 at
0.841 of $N$, against the window's 0.149. The pre-registered rules are at odds here,
not the replay: D3 admits a depth whose gap is up to $\varepsilon = 0.01$, while the family
comparison uses the band $\tau = 0.005$. We report the registered result and do not re-plan.
Over all 31 primary cells we replayed (MS MARCO and NQ), the lexical union wins 6
(P11 at 50, P11 at 25, P2 at 25, P9 at 25, N3 at 25, P5p at 50; N3 is P3's model pair on NQ, below), buying 0.0051--0.0134 of quality for 0.84--0.91 of
$N$ paid before service, 3.2--5.7$\times$ the window's cost where the window
reached target quality (it did not in 3). The committed planner omits it because its registered statistic prices only S2's windows, not
because it never wins. Counting each strategy's full candidate union from the serving side
afterwards gives S3--S5 a conservative price (it bounds inline work; it is not first-arrival
cost); with it, the static winner matches the replay at $K'{=}50$ on all ten text pairs at seed 0
and on 34 of the 36 new Zipf replays, missing P5 and P14 at $\alpha = 1.2$. At the depths the margin-corrected D3 chooses, it wins at seed 0 on P11 alone
among the MS MARCO pairs{} (seeds: Section~\ref{sec:results}). A deployment that values that
quality at a cost near $N$ should add it.P11, which D3 kept at 50 because $G(25) = 0.0137> \varepsilon$, was also replayed
at 25 for information: there the window's debt is 0.0137, as $G$ said. (The
12 of 13 above scores the replays at 50, where P2 and P9 match and P11 misses.)

\begin{table}[htbp]
\centering
\caption{D3's $K'{=}25$ decisions, committed before these replays. Adv.: S2's debt minus S4's.
Verdict: cost / quality. $^\dagger$Never at target quality.}
\label{tab:d3}
\footnotesize\setlength{\tabcolsep}{2.6pt}
\begin{tabular}{@{}lrrrrrll@{}}
\toprule
Pair & $G(25)$ & $\cov(25)$ & cost & debt & adv. & winner & verdict \\
\midrule
P2 & $+0.0094$ & $0.1486$ & never$^\dagger$ & $+0.0102$ & $+0.0105$ & S4 & -- / \ding{51} \\
P3 & $-0.0032$ & $0.1481$ & $0.1480$ & $-0.0014$ & $+0.0007$ & S2 & \ding{51} / \ding{51} \\
P4 & $-0.0011$ & $0.1452$ & $0.1451$ & $-0.0001$ & $+0.0029$ & S2 & \ding{51} / \ding{51} \\
P5 & $-0.0088$ & $0.1452$ & $0.1451$ & $-0.0038$ & $+0.0001$ & S2 & \ding{51} / \ding{55} \\
P9 & $+0.0050$ & $0.1486$ & $0.1485$ & $+0.0063$ & $+0.0073$ & S4 & \ding{51} / \ding{51} \\
P10 & $-0.0002$ & $0.1486$ & $0.1485$ & $+0.0036$ & $+0.0046$ & S2 & \ding{51} / \ding{51} \\
P12 & $+0.0014$ & $0.1452$ & $0.1451$ & $+0.0015$ & $+0.0032$ & S2 & \ding{51} / \ding{51} \\
P13 & $-0.0018$ & $0.1481$ & $0.1480$ & $-0.0004$ & $-0.0005$ & S2 & \ding{51} / \ding{51} \\
\bottomrule
\end{tabular}
\end{table}

\textbf{How much target encoding D1 needs.} D1 reads $M_T$, which needs a target-encoded
sample. Re-measured on 1, 5 and 25\% of the corpus (judged passages kept, three seeds), a sample
understates the gap in 100 of 108 draws, since a smaller corpus has fewer
distractors. At 25\% D1 decides as on the full corpus, on every seed, for 10 of
12 pairs; the misses, P3 and P13, go to a rebuild, which costs quality, not
encoding. The error is not one-sided in principle: on P3 the $10^6$ gap overstates the full
collection's (Section~\ref{sec:scale}).

\textbf{A prediction on a corpus the planner has not seen.} On the BEIR~\cite{beir} NQ~\cite{nq} subset
($10^5$ Wikipedia passages, 1,000 judged queries) we run the same ten model pairs plus an identity control, eleven in all, compute the
plan from static inputs, commit it, and only then replay each pair at the depth the plan names.
NQ pairs keep their MS MARCO numbers with N for P (N2 is P2's model pair); N0 is the control. The plan named 5 rebuilds and 6 windows. Replayed, the family held
on 10 of 11 and every window cost exactly its planned coverage. Quality held on 4 of 6, with deviations on both sides of $G$. The miss, N14 (gap +0.0043), arises because the replay
anchors its band at a lexical union's debt ($-0.0020$), which D2 excludes; among
admitted families the rebuild wins as planned.

\subsection{Why no selector is needed}\label{sec:res-select}
The study was designed around a regime map: two cheap compatibility measurements, $\kappa$ and
$\rho$, predicting the winning strategy. We pre-registered five hypotheses with falsification
conditions (Table~\ref{tab:hyp}). Three fail, and H2 holds by only 0.00006 of regret, about 1.1\% of
$\tau$. So the planner is three rules from the frame, not a fitted model: the data leave little to learn.

\begin{table}[htbp]
\centering
\caption{Pre-registered hypotheses and outcomes.}
\label{tab:hyp}
\footnotesize\setlength{\tabcolsep}{2.5pt}
\begin{tabular}{@{}lp{0.30\columnwidth}p{0.42\columnwidth}l@{}}
\toprule
& Hypothesis & Falsified if \ldots{} measured & Outcome \\
\midrule
H1 & $G(K')$ rises as $\kappa$ falls & $r_s(-\kappa,G) < 0.5$ at $K'{=}50$; $+0.17$ (11 pairs) & \textbf{fals.} \\
H2 & $(\kappa,\rho)$ beats majority in leave-one-out regret & regret $\ge$ majority's; 0.00079 vs 0.00085& holds \\
H3 & low-$\kappa$/low-$\rho$ cell non-empty & exploratory; P9, P11, but $r_s(\kappa,\rho) = -0.52$ & holds \\
H4 & a probe at $\le 2\%$ of $N$ gets $\rho$ within $\pm0.05$ & $>1$ pair outside; 2 of 10 & \textbf{fals.} \\
H5 & signal order beats rand & S2: 7 better, 6 worse ($p = 1.00$); S2/S4/S5: 8, 20 ($p = 0.036$) & \textbf{fals.} \\
\bottomrule
\end{tabular}
\end{table}

\textbf{The winner is almost constant across pairs.} At the primary setting the 13 winners
are S2 9 times, S0 3 times (the D1 cases) and S4 once. Within the window
family the variants tie: 8 pairs report a winner chosen from 3--4 combinations
with identical debt, picked by the simplicity order, and S2's orders tie within
$\tau$ in 45 of 45 seeded replays. The ties belong to the operating point: at
$\mathrm{tr}{=}64$ the background has covered 0.03 of the corpus when the
population is served, against 0.63 at $\mathrm{tr}{=}4$, where order does move
debt.

\textbf{No feature does measurably better than a constant.} In leave-one-pair-out prediction,
1 of 6 feature rules (1-NN on $(\kappa,\rho)$ at $K'{=}50$) beats the
majority rule, which uses no features, and only by 0.00006 of regret
(1.1\% of $\tau$), with equal accuracy. The majority rule's largest regret,
0.00509, is on P11, where the replay selects S4, which D2 excludes.

\begin{figure}[t]
\centering
\includegraphics[width=0.8\columnwidth]{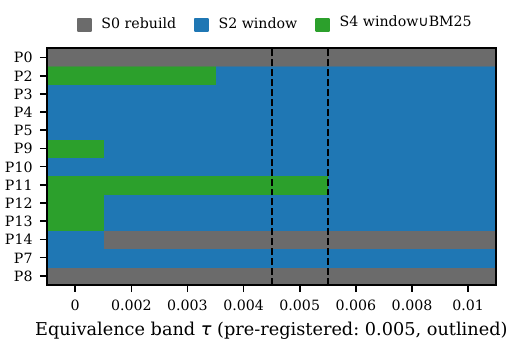}
\caption{Winner family by pair as $\tau$ varies, primary traffic intensity.}
\label{fig:tau}
\end{figure}

\textbf{Which pairs deviate depends on $\tau$} (Fig.~\ref{fig:tau}).
Sweeping the winner rule over every pair and traffic intensity (52 configurations, each
at 8 values of $\tau \in [0, 0.01]$) never selects S1, S3 or S5. S2 is selected in
289 cells, S0 in 83 and S4 in 44. One step either side of the registered
$\tau$ changes 1 and 4 cells. One or two pairs always deviate from S2; which
ones is a function of $\tau$, which we did not retune.

\textbf{The cheap probe and the ordering signal do not rescue a selector.} H4: a stratified
probe cuts the error of estimating $\rho$ 3.6$\times$ at 1.2\% of $N$ but
misses the bound on 2 pairs, and $\rho$ does not predict the winner anyway. H5: the
signal is real (AUC 0.622--0.650 on text), but every effect of ordering by it is
inside $\tau$ (Table~\ref{tab:hyp}), almost all at low traffic intensity; only S3, never
selected, gains outside the band, so the pooled family omits it.

\subsection{The regime where reuse cannot help}\label{sec:res-regime}
Windowed reuse defers $1-\cov(K')$ of the work, and $\cov(K') \to 1$ as $|Q|K'$ approaches $N$.
Our code corpus is past that point: 43,730 servable queries over 45,773 functions.
At $K'{=}25$ the windows already touch all but 12 functions
(coverage 0.99974), and from $K'{=}50$ upward they touch every one. The ceiling on
deferral is zero however compatible the models are, a property of corpus and workload that the
planner reads from the same retrieval pass. A window then costs $N$ and a rebuild is as cheap,
so selection is vacuous and the code pairs are no evidence.

\section{Scale}\label{sec:scale}
Our replays use $10^6$ passages, two to three orders of magnitude below the deployments that
motivate this work. The union of windows grows with the number of \emph{queries} far more than with
the corpus (Table~\ref{tab:scale}): at a fixed workload and depth, $\cov(K') \le \min(1, |Q|K'/N)$, so the deferrable share
$1-\cov(K')$ has a floor that rises with $N$. At our operating point $|Q|K'/N = 0.349$; the
measured ceiling is 0.7357 and the closed form gives 0.7054.

Two effects could still break the planner: a fixed $K'$ is a shrinking share of a growing corpus,
so $G(K')$ should rise and D3 answer with a deeper window; and the gap itself may move. We measured
both on the full collection (Table~\ref{tab:scale}). Before any full-corpus vector existed we
committed four predictions per pair (\texttt{PREREG\_SCALE.md}): neither (a)~the gap nor (b)~$G$
shrinks, (c)~window documents do not shrink, and (d)~coverage stays below the bound.

\begin{table}[t]
\centering
\caption{Planner inputs at $10^6$ and at all 8,841,823 passages, same queries; docs and $\cov$ at
$K'{=}50$. Failed predictions (Section~\ref{sec:scale}): (a)~gap, (b)~$G$, (c)~window documents, (d)~coverage bound.}
\label{tab:scale}
\footnotesize\setlength{\tabcolsep}{2.2pt}
\begin{tabular}{@{}llrrrrrrl@{}}
\toprule
Pair & $N$ & gap & $G(25)$ & $G(50)$ & D3 & docs & $\cov$ & failed \\
\midrule
P2f & 1.0M & $+0.0471$ & $+0.0094$ & $+0.0022$ & 25 & 264,274 & 0.2643 &  \\
 & 8.8M & $+0.0644$ & $+0.0223$ & $+0.0078$ & 50 & 331,388 & 0.0375 & none \\
\midrule
P3f & 1.0M & $+0.0154$ & $-0.0032$ & $-0.0029$ & 25 & 261,214 & 0.2612 &  \\
 & 8.8M & $+0.0099$ & $+0.0016$ & $-0.0037$ & 25 & 332,874 & 0.0376 & (a), (b) \\
\midrule
P9f & 1.0M & $+0.0317$ & $+0.0050$ & $-0.0010$ & 25 & 264,274 & 0.2643 &  \\
 & 8.8M & $+0.0545$ & $+0.0175$ & $+0.0081$ & 50 & 331,388 & 0.0375 & none \\
\bottomrule
\end{tabular}
\end{table}

Of these 12 static predictions (three pairs, four each), 10 hold, including both
coverage predictions. From $10^6$ to 8,841,823 passages the documents in the $K'{=}50$ windows grow by only
25--27\%, so $\cov(50)$ falls 6.94--7.05-fold, within the 8.84 that (c) allows, to
0.0375--0.0376 (bound 0.0387), and the most a window can defer rises to 96.2--96.3\%. The failures, (a) and (b) on P3f, are on the quality side: the gap shrinks on P3f (0.0154 to 0.0099), and $G(50)$ with it. D1
decides as at $10^6$ on 3 of 3; at $\varepsilon = 0.01$, D3 deepens P2f (to 50) and P9f (to 50), whose $G(25)$ now exceeds
$\varepsilon$, and keeps P3f, whose $G(25)$ does not.
A smaller corpus need not bound the quality inputs: the samples of
Section~\ref{sec:res-plan} mostly understate the gap, while P3's $10^6$ gap overstates it. They
must be measured on the index that will serve, where coverage needs no target encoding and $G$
only the judged queries' windows. \textbf{Replayed at 2M, 4M and 8.8M.} We cut nested 2M and 4M corpora from the full collection and
replayed P2f, P3f and P9f at all three sizes at a fixed $K'{=}50$, which is D3's 8.8M choice for
P2f and P9f; the 2M and 8.8M predictions were committed from the static sweeps before their
replays (six replay predictions, separate from the static ones), and 4M is supporting. Cost held: in all nine replays the window's cost-at-service is
within 100 documents of the static coverage of the served windows, which falls from 0.26 at
$10^6$ to 0.148, 0.079 and 0.0375. Both pre-registered sizes score 3 of 3 on cost, quality and
family: S2 wins every pair at every size, and at 8.8M its debts, $+0.0090$, $-0.0040$ and
$+0.0092$, sit inside the pre-registered band $[G(50)-\tau,\,G(50)+m]$, $m = 0.006$ the margin of
Section~\ref{sec:res-plan}, around $G(50) = +0.0078$,
$-0.0037$ and $+0.0081$:
measured at the serving size, the quality inputs predicted the replay.

\section{Related Work}\label{sec:related}
\textbf{Embedding model upgrades.} Work on upgrades falls into reuse mechanisms, compatibility
measures and training-side methods. Drift-Adapter maps new queries into the legacy space and
reports 95--99\% of re-embedding Recall@10 at under $10\,\mu$s per query~\cite{drift-adapter};
our S1/S3 are its family, and we add the bar a migration decision uses, $M_S$. EmbedFlow serves
a target model over source-index candidates while materializing target vectors~\cite{embedflow};
S2 is its mechanism and $G(K')$ its diagnostic. We add the cost model, the decision whether to
migrate at all, and the comparison across strategies. Residual reachability reuses old graph structure to build the new index~\cite{dgm}. Our
compatibility axis follows Procrustes alignment~\cite{procrustes}, CKA~\cite{cka} and relative
representations~\cite{relrep}. Training-side methods make a new model backward-compatible with
the old index~\cite{bct,hyperbolic-bct} or map an updated model's queries into the old
space~\cite{qdc}, but they need control over training; we study upgrades between released models,
whose relative quality benchmarks report~\cite{mteb}.

\textbf{Dynamic vector indexes and stale representations.} The cost model reads only the
serving index's top-$K'$ windows, which graph, partition and quantization
indexes~\cite{faiss,diskann,scann,graph-ann-survey} all return. Updatable indexes~\cite{spfresh,freshdiskann,quake} assume the new vector exists, which is
the cost studied here. S4 and S5 combine dense and lexical candidates, as hybrid search
does~\cite{deg}. Query-time recomputation~\cite{leann} and corrector
networks~\cite{corrector} reduce re-encoding in other settings. CIPBench finds,
under a fixed encoder, that an exposure signal adds nothing over change history for choosing
what to re-embed~\cite{cipbench}, consistent with our ordering result.

\textbf{Systems.} Vector databases~\cite{milvus,vdbms-survey} treat the embedding model as fixed;
the planner borrows System-R's stance~\cite{selinger} of estimating from statistics, and learned
indexes face a related question of when to replace a stale model~\cite{alex}.

\section{Discussion and Conclusion}\label{sec:disc}
\textbf{Recommendations.} \emph{Quality first:} measure $M_T - M_S$ on a judged
sample of the index that will serve (a smaller corpus can misstate it either way); if it is within the band you declare in advance (the largest
regression you would not notice), the upgrade buys nothing distinguishable: rebuild without reuse, or skip it.
\emph{Then cost:} measure $\cov(K')$ on the serving index (one retrieval pass, no target
encoding). If $1-\cov(K')$ is small, rebuild. Otherwise use plain window re-ranking at the
smallest depth whose $G$ fits your quality band minus the margin by which served debt exceeds
$G$ (0.006 here; Section~\ref{sec:res-plan}), with any completing order, if a cold query can wait one batched forward
pass. Plain S2 is the cost-efficient default, unless the lexical union's gain is worth paying
3.2--5.7$\times$ as much up front (Section~\ref{sec:res-plan}).

\textbf{Limitations.} D3's depths were scored at one traffic intensity, on MS MARCO and NQ, and its
quality criterion is tight (P5 misses by 0.00006; debt exceeds $G$ by up to
0.0058). The scale replays cover three pairs at one depth, $K'{=}50$. At Zipf(1.2) the static scan leaves
18 effective judged queries and the replayed arrival order about 12, and D1 decides near its threshold unstably. D2 excludes
S3 whenever the bridge falls short of $\tau$, which can miss a winning S3, as on P5r. Latency is measured on an idle accelerator; a shared
host multiplied it by up to 3. Table~\ref{tab:claims} lists what is not covered.

\textbf{Conclusion.} Once a migration must finish, every strategy encodes the same $N$ documents; strategies differ
only in when that work is paid and in the quality served meanwhile. Window cost can be read from
the serving index before migrating, and pre-registered replays at 8,841,823
passages matched it on all three pairs tested; a bridge must beat the old model; and quality inputs, skew included, must be measured on the serving index.
Code and experimental artifacts are available at \url{https://github.com/nnzhaocs/ReuseRAG}

\bibliographystyle{IEEEtran}
\bibliography{references}

\end{document}